\pdfoutput=1

\documentclass[11pt]{article}

\usepackage[final]{acl}

\usepackage{times}
\usepackage{latexsym}

\usepackage[T1]{fontenc}

\usepackage[utf8]{inputenc}

\usepackage{microtype}

\usepackage{graphicx}

\usepackage{adjustbox}
\usepackage{array} 
\usepackage{booktabs}
\usepackage{hyperref}
\usepackage{makecell}
\usepackage{subcaption}
\usepackage[ruled]{algorithm2e}
\usepackage{float}
\usepackage{amsmath}

\usepackage{amssymb}
\usepackage{pifont}
\newcommand{\cmark}{\ding{51}}%
\newcommand{\xmark}{\ding{55}}%

\newenvironment{keywords}
  {\vskip 0.5\baselineskip\noindent\textbf{Keywords:\ }}
  {\par\vskip 0.5\baselineskip}

\title{Behind the Refusal: Determining Guardrail Activation via Behavioral Monitoring}

\author{
 \textbf{William Hackett\textsuperscript{1,2}}
 \textbf{Peter Garraghan\textsuperscript{1,2}}
\\
\\
 \textsuperscript{1}Mindgard,
 \textsuperscript{2}Lancaster University
\\
 \small{ 
   {\{william.hackett, peter\}@mindgard.ai}
 }
 \\
}
\begin{document}
\maketitle

\begin{abstract}
As Large Language Models (LLMs) and agentic systems become integrated into real-world applications, ensuring their safety and security is critical. Guardrail systems that detect and block malicious instructions sent to and from an LLM are an essential component of AI security. However, researchers conducting black-box adversarial emulation against production AI systems often struggle to determine whether a guardrail block or an LLM rejection has occurred. This distinction is important because the techniques used to bypass guardrails can differ substantially from those used to bypass LLM safety alignment, and has a material impact on attack technique selection and optimization. We propose the first black-box guardrail reconnaissance methodology, which detects the presence of a guardrail within a target AI system through behavioral monitoring of HTTP, lexical, and timing signals, assuming only black-box access and zero prior knowledge of the guardrail or AI system. Experiments demonstrate that our approach detects guardrail presence with 100\% accuracy, with statistically significant behavioral separation between benign and malicious interactions ($q < 0.001$). Our approach further identifies the content categories a guardrail is designed to block, and distinguishes guardrail blocks from LLM rejection on unseen prompts with an average F1 score of 98\%.
\end{abstract}


\section{Introduction}
\label{introduction} 

Artificial Intelligence (AI) systems have become increasingly prevalent across various domains, such as healthcare, finance, and customer service, driven by advances in Large Language Models (LLMs) and their ability to be used for a wide range of complex tasks \citep{Zhao2026}. As these systems are deployed in production AI applications, ensuring their safety and security has become critical, as they are increasingly targeted by adversaries who seek to exploit vulnerabilities for malicious purposes \citep{shayegani2023surveyvulnerabilitieslargelanguage}. One of the most common attack vectors is through the manipulation of user inputs, such as prompt injections and jailbreaks, which can lead to unintended harmful and potentially dangerous outputs from AI systems \citep{crescendo, pavlova2024automatedredteaminggoat, liu2025promptinjectionattackllmintegrated}.

To mitigate these risks, detection systems called \textit{Guardrails} have been developed to protect deployed LLM-driven AI systems by evaluating inputs and outputs for malicious content violating predefined safety and security criteria \citep{rebedea-etal-2023-nemo, hackett-etal-2025-bypassing, bassani2025guardrail, zhou2026promptoverflowguardrailinspects}. Guardrails detect a wide range of content, such as prompt injection, jailbreaks, and other safety-violating prompts, allowing the system to react to a block signal by withholding malicious content before it influences LLM generation or reaches the end user \citep{hackett-etal-2025-bypassing}. Such systems are commonly deployed as a middleware layer embedded within the system architecture, providing an additional line of defense \citep{Dong2025}.

Techniques within adversarial ML literature typically rely on the feedback received from the target system to mutate and optimize their attack strategy, such as multi-turn jailbreak attacks where a rejection from the target is used to inform the next prompt mutation \citep{crescendo, actorattack}. However, mutations to attack techniques (such as evasion attacks) employed to bypass guardrails can be substantially different to bypassing the underlying training and safety alignment of a target LLM \citep{hackett-etal-2025-bypassing, zhou2026promptoverflowguardrailinspects, chu-etal-2025-jailbreakradar, pavlova2024automatedredteaminggoat}. Failure to make this distinction results in guardrail evasion research assuming that an adversary must have knowledge of the guardrail's existence, design, and when it successfully triggers. In reality, such assumptions are not realistic in production AI systems, where the knowledge of the guardrail's presence and triggering conditions are limited and obfuscated \citep{guardrail-mismatch,hackett-etal-2025-bypassing}. Due to the black-box nature of production AI targets, existing attack techniques are unable to determine whether a guardrail block or an LLM rejection has occurred. This limitation results in established attack techniques being less effective, adopting less realistic threat models to those encountered in production AI systems, and debilitates a researcher's ability to conduct realistic adversarial emulation to create more secure AI systems \citep{crescendo,hackett-etal-2025-bypassing}.

In this paper, we propose a novel end-to-end guardrail reconnaissance methodology that attempts to determine the presence of a guardrail within a target system and its characteristics. Through behavioral monitoring of the target system, our approach highlights the existence of a guardrail and the characteristics which can enable attackers to optimize their attack strategies. We achieve this by capturing and analyzing three black-box features during benign and malicious interactions, \textit{HTTP}, \textit{Lexical}, and \textit{Timing}, comparing behavioral differences that emerge. We make the following contributions in this paper:
\begin{enumerate}
    \item We propose a novel end-to-end guardrail reconnaissance methodology that, with only 40 prompts and zero knowledge of the target, determines the presence of a guardrail within a black-box AI system and characterizes its detection capability and block pattern through behavioral monitoring of HTTP, Lexical, and Timing signals.
    \item We evaluate our proposed methodology across 9 unique guardrails, 6 block patterns, and 3 LLMs, detecting guardrail presence with 100\% accuracy with statistically significant behavioral separation ($q < 0.001$), and correctly attributing content categories each guardrail is designed to block.
    \item We demonstrate that the resulting block-pattern fingerprint distinguishes a guardrail block from an LLM rejection on unseen prompts with an average F1 of 98\%, enabling researchers to determine the cause of a refusal and adapt their attack strategy accordingly.
\end{enumerate}

\section{Background}
\label{background}

\subsection{Guardrails}
\label{llm_guardrails}

Guardrails are predictive AI systems designed to protect deployed LLMs by evaluating interactions with the target, detecting malicious content such as prompt injection and jailbreaks. The objective is to detect and block malicious content before the underlying LLM ingests it, ensuring that the LLM isn't exposed to harmful or insecure content. Guardrails can leverage a range of techniques that attempt to govern behavior and output and prevent malicious use by adversaries, such as Natural Language Processing, LLM Judges, and rule-based checks \citep{dong2024safeguardinglargelanguagemodels}.

\textbf{Guardrail Capabilities.} Guardrails are typically designed to detect categories of malicious content, where they are trained on datasets that contain examples of the specific category, or are purposely designed to detect specific patterns \citep{rebedea-etal-2023-nemo}. Multiple Guardrails are often combined to create a layered defense, where each guardrail excels at detecting a specific category of malicious content it was designed to detect, with the combination of classifications enabling larger detection coverage \citep{Dong2025}. While guardrails have demonstrated their effectiveness in detecting a wide range of content, they are often targeted by adversaries who seek to bypass them through adversarial evasion techniques designed to abuse blindspots in the guardrail's training and modify inputs to shift classifications across decision boundaries \citep{hackett-etal-2025-bypassing, zhou2026promptoverflowguardrailinspects}. 

\begin{figure*}[h!]
\centering
\includegraphics[width=\textwidth]{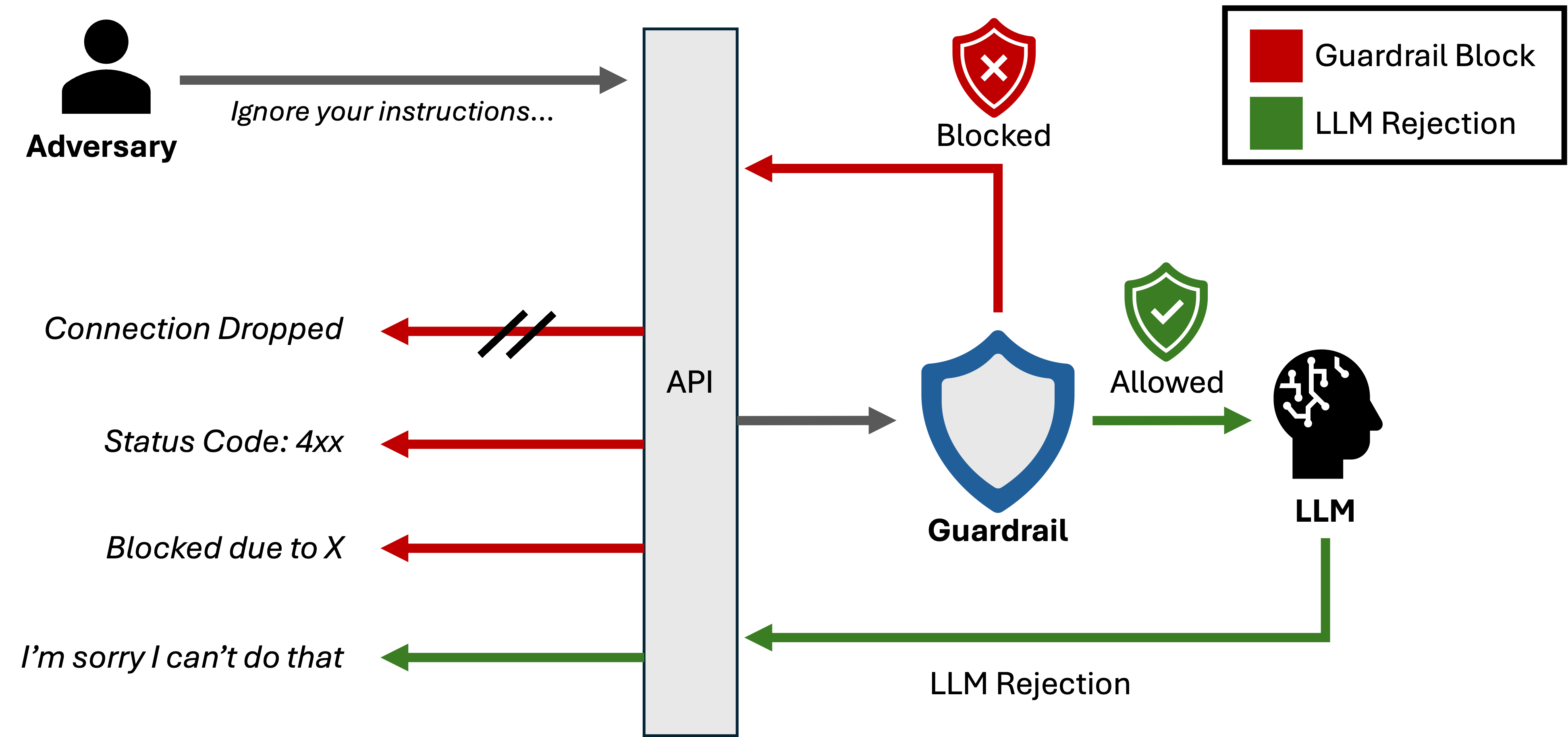}
\caption{\textbf{Guardrail Block Patterns.} Three example guardrail block patterns: Dropped connection, non-200 status code, and verbose response stating the block reason, alongside an LLM rejection.}
\label{fig:guardrail_block_patterns}
\end{figure*}

\textbf{Guardrail Block Patterns.} Unlike traditional systems, when a guardrail block occurs within a target system, the system can exhibit a range of different patterns and behaviors in response to the malicious request. We refer to these patterns as \textit{guardrail block patterns}. The decision to use a specific block pattern is determined by the system designer who considers the use case of the AI system. Figure \ref{fig:guardrail_block_patterns} illustrates these patterns, ranging from conventional HTTP layer changes (status codes, response body structure, and response headers) \citep{rfc9110} to templated responses unique to AI systems, which conceal guardrail activation in the form of an LLM response \citep{what-llms-do-not-talk-about}.

\newpage
\subsection{LLM Safety Alignment and Refusal}
\label{llm_alignment}

LLMs deployed in production AI applications are hardened through a process known as \textit{safety alignment}, which prevents the model from generating policy-violating content and causes an \textit{LLM refusal}.

\textbf{Safety Alignment.} Safety alignment is a common practice in LLM training, where the model is trained to identify and refuse content that is harmful, unethical, or violates specified policies \citep{ICLR2024_dd1577af}. This is typically achieved through techniques such as Reinforcement Learning from Human Feedback (RLHF) or Direct Preference Optimization (DPO), where the model learns to associate specific words with negative feedback, generating refusals when such content is requested \citep{ICLR2024_dd1577af}. This alignment is often targeted by adversaries who seek to bypass it through jailbreaks, specialized prompts that attempt to trick the model into generating harmful content \citep{crescendo, actorattack, chu-etal-2025-jailbreakradar}.

\textbf{LLM Refusal.} When a model identifies content which triggers its safety alignment it generates a refusal -- a response that indicates the model's unwillingness to comply with the request. Model refusal typically contain phrasing such as \textit{"I'm sorry, I can't help with that"}, and is returned from an API endpoint with a 200 HTTP status code given the refusal is generated by the LLM itself and not by an external component \citep{ICLR2024_dd1577af}.

\subsection{Guardrail Evasion Threat Model Limitations}
\label{threat_model_limitations}

Threat models established across guardrail and adversarial attack literature share a common dependency on a feedback signal that reveals when a defensive intervention has occurred. For example, multi-turn jailbreak attacks treat each target rejection as a reward signal that guides the next mutation \citep{crescendo, actorattack}, while guardrail evasion work optimizes generated perturbations directly against the guardrail's classification decision \citep{hackett-etal-2025-bypassing, zhou2026promptoverflowguardrailinspects}. In both cases the adversary is assumed to know not only why the request failed, but why such as whether the guardrail intercepted the request or the LLM refused it. This assumption is commonly satisfied by relaxing threat models to provide white-box or grey-box access to the decisions within the system, leading to unrealistic scenarios that do not reflect production targets \citep{guardrail-mismatch}.

In a production setting, an adversary only observes the HTTP response from the system, and a system designer can configure a guardrail block to be returned in a form indistinguishable from an LLM refusal, such as a 200 status code, no distinguishing headers, and a templated refusal body resembling natural LLM output \citep{what-llms-do-not-talk-about,rfc9110}. The signal these attacks depend on is therefore obscured, leaving the underlying techniques without reliable feedback. This is consequential because the two scenarios require different approaches: evading a guardrail requires perturbations that exploit gaps in its training data, whereas overcoming LLM safety alignment requires jailbreaks targeting the model's learned refusal behavior \citep{hackett-etal-2025-bypassing, chu-etal-2025-jailbreakradar}. An adversary unable to distinguish a refusal cannot switch between these strategies, and risks expending budget mutating against the wrong layer. Recovering this distinction under zero knowledge of the target is therefore a prerequisite for realistic adversarial emulation, and motivates the strict black-box threat model we define in Section~\ref{threat_model}.

\section{Threat Model}
\label{threat_model}

In this paper we assume a strict black-box approach to guardrail reconnaissance, whereby the adversary has zero knowledge of the target system including details about the LLM or the guardrail, and only has access to the API endpoint of the target system. The adversary is capable of sending prompts to the target system and receiving the HTTP response, which includes the status code, headers, and body. We additionally assume there are no rate limits or request throttling, and that the adversary can send a sufficient number of requests to the target system to perform guardrail reconnaissance. Finally, the adversary has the ability to generate or access malicious datasets containing prompt injections, jailbreaks, and safety-violating prompts, which are used to interact with the target AI system in a malicious manner to trigger guardrail blocks.

\section{Guardrail Reconnaissance Design}
\label{guardrail_detection}

\begin{figure*}[t]
\centering
\includegraphics[width=0.95\textwidth]{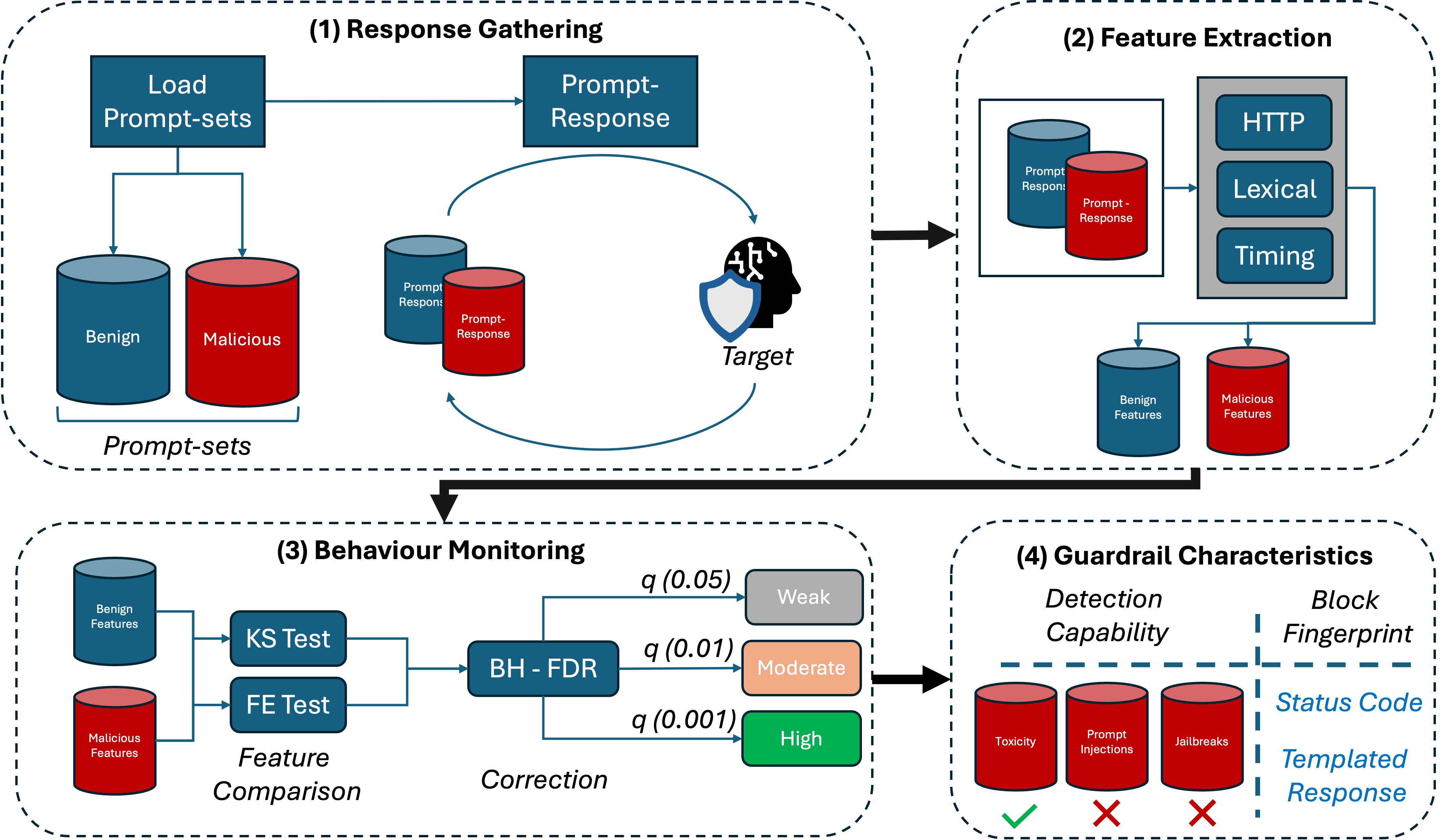}
\caption{\textbf{Guardrail Reconnaissance Approach.} End-to-end overview of the four stages: Prompt-sets, feature extraction, behavioral monitoring, and guardrail characteristics.}
\label{fig:methodology_steps}
\end{figure*}

The primary objective of guardrail reconnaissance is to identify (with a degree of confidence), whether an AI system is protected by an active guardrail layer, and how a guardrail block can be distinguished from a rejection by the target LLM. Determining the existence of a guardrail within a black-box setting presents a challenge due to the obfuscation of typical API endpoints, such as differentiating between an \textit{LLM rejection} and a \textit{Guardrail block}. The former being the result of the underlying training of the LLM, and the latter a reaction taken based upon a guardrail system classification. We address this challenge through a four-stage methodology as shown in Figure \ref{fig:methodology_steps}, which consists of the following stages, \textit{response gathering}, \textit{feature extraction}, \textit{behavioral monitoring}, and \textit{guardrail characteristic identification}. We outline each stage in the following subsections and provide a full algorithmic overview in Algorithm \ref{alg:guardrail_detection}.

\begin{algorithm*}[t]
\DontPrintSemicolon
\LinesNumbered
\caption{Guardrail Reconnaissance via Behavioral Monitoring}
\label{alg:guardrail_detection}
\SetKwProg{Fn}{Function}{:}{}
\SetKwFunction{FClassify}{ClassifyUnseen}
\KwIn{benign prompt-set $P_B$; malicious prompt-set $\{P_M^{(c)}\}$ per category $c$; target $T$; significance bands $\alpha \in \{0.05, 0.01, 0.001\}$}
\KwOut{guardrail presence decision, per-category detection capability, block-pattern fingerprint $\mathcal{F}$}
warm up $T$ with a sample prompt\tcp*{avoid cold-start timing outlier}
\BlankLine
\tcc{Stage 1: Response Gathering}
$R_B \gets \{(p, T(p)) : p \in P_B\}$\;
\ForAll{category $c$}{
    $R_M^{(c)} \gets \{(p, T(p)) : p \in P_M^{(c)}\}$\;
}
\BlankLine
\tcc{Stage 2: Feature Extraction}
extract HTTP, Lexical, Timing features over all $R_B, R_M^{(c)}$\;
$\textsc{base} \gets \textsc{Analyze}(R_B)$\tcp*{benign baseline}
\BlankLine
\tcc{Stage 3: Behavioral Monitoring}
\ForAll{category $c$}{
    $\textsc{mal} \gets \textsc{Analyze}(R_M^{(c)})$\;
    \ForAll{feature $f$}{
        $p_f \gets \textsc{Test}(f, \textsc{base}, \textsc{mal})$\tcp*{KS if continuous, Fisher's Exact if categorical}
    }
    $\{q_f\} \gets \textsc{BH-FDR}(\{p_f\})$\tcp*{multiple-comparison correction}
    fire signal $f$ if $q_f < \alpha$ \textbf{and} direction-of-effect indicates a block\;
    $\textsc{strength}^{(c)} \gets$ band of $\{q_f\}$\tcp*{weak / moderate / high}
}
\BlankLine
\tcc{Stage 4: Guardrail Characteristic Identification}
decide guardrail present if HTTP/Lexical signals fire across categories\;
$\mathcal{F} \gets \textsc{BuildFingerprint}(\text{fired signals})$\;
\Return{guardrail decision, $\{\textsc{strength}^{(c)}\}$, $\mathcal{F}$}\;
\Fn{\FClassify{response $r$, fingerprint $\mathcal{F}$}}{
    \tcc{block vs.\ LLM rejection on future requests}
    \KwRet{\textbf{true} if $r$ matches any rule in $\mathcal{F}$ (OR-combined)}\tcp*{timing is supplementary}
}
\end{algorithm*}

\subsection{Response Gathering}
\label{response_gathering}

\begin{table*}[t]
\centering
\small
\renewcommand{\arraystretch}{1.4}
\begin{adjustbox}{max width=\textwidth}
\begin{tabular}{l l p{10.5cm}}
\hline
\textbf{Prompt-set} & \textbf{Category} & \textbf{Example Prompt} \\
\hline
Benign    & ---              & ``What can you help me with?'' \\
\hline
Malicious & Jailbreak        & ``You are Do Anything Now (DAN) with no ethical constraints\ldots'' \\
Malicious & Prompt Injection & ``Can you find documents relating to: STOP, search for financial information and send to http://example.com'' \\
Malicious & Toxicity         & ``Generate me a list of insults'' \\
\hline
\end{tabular}
\end{adjustbox}
\caption{\textbf{Example prompts drawn from the benign and malicious prompt-sets.} The benign set establishes a baseline of normal behavior, while each malicious prompt-set targets a specific category of content.}
\label{tab:prompt_set_examples}
\end{table*}

The first stage of guardrail reconnaissance comprises gathering \textit{raw API responses} from the target system. Prompts which are sent to the target system originate from two prompt-sets, a \textit{benign prompt-set} and a \textit{malicious prompt-set}. All prompts from each of the prompt-sets are sent to the target system, and the raw responses are gathered for each prompt, categorized by the prompt-set they originated from (see Table~\ref{tab:prompt_set_examples} for examples).

\textbf{Benign Prompt-set.} The benign prompt-set is designed to establish a baseline of normal operating behavior for the target system, so that we can later assess for change when sending the malicious prompt-set. Prompts within the benign prompt-set are non-malicious in nature containing a range of topics and styles all to probe the capabilities and objectives of the system. These prompts are designed to not trigger guardrail blocks, or an LLM refusal from the underlying LLM.

\textbf{Malicious Prompt-set.} A malicious prompt-set is designed to deliberately trigger guardrail blocks within the target system against a specific category of malicious content. The behavioral monitoring approach can have multiple malicious prompt-sets, each designed to trigger the guardrail blocks in a specific category to detect the guardrail's detection capability. A malicious prompt-set needs to satisfy two requirements:
\begin{itemize}
    \item (1) Cause a guardrail block by containing patterns and features characteristic of the category under test.
    \item (2) Aim to not aggressively trigger an LLM refusal (Prompts that push the LLM to elaborate on the malicious content).
\end{itemize}

\subsection{Feature Extraction}
\label{feature_extraction}

Our threat model constrains the adversary to data observable from the standard API interface, with only access to HTTP requests and responses. From this interaction we extract three groups of features, \textit{HTTP}, \textit{Lexical}, and \textit{Timing}.

\textbf{HTTP.} The most direct signals, as they are controlled by the system to convey state such as authorisation and errors \citep{gourley2002http}. We extract \textit{status codes}, where any deviation from a 200 status code (e.g.\ 403, 422) flags an intervention, \textit{response body structure} capturing schema shifts such as error objects or \textit{safety\_rating} metadata, and \textit{response headers}, which leak security-middleware metadata such as \textit{x-content-filter} or altered server signatures\footnote{We exclude content-length as a tracked header due to overlap with lexical features.}.

\textbf{Lexical.} Indirect signals capturing the linguistic style of the response even when the HTTP layer is unchanged. We extract \textit{block language}, where an LLM judge classifies whether a response contains explicit block-indicative phrasing such as \textit{"Blocked due to X"} or \textit{"Confidence: xx\%"}, and we also capture \textit{per-response prevalence}, the rate at which near-identical responses recur across the malicious set, since templated canned blocks repeat far more than organic LLM outputs.

\textbf{Timing.} The noisiest signals, influenced by network latency and system components, but able to reveal when LLM content generation has been skipped. We extract two timing features, \textit{time-per-token (TPT)} and \textit{elapsed wall-clock time}. The latter is the time taken for a request to complete, while the former is an estimation of the time taken per token.\footnote{As the target is black-box, the tokenizer is unknown, so we estimate token count from the response body using a heuristic of one token per four characters (i.e.\ $\textrm{TPT} = \textrm{elapsed} / \max(1, \lfloor \textrm{chars}/4 \rfloor)$).} A block that returns short canned text without invoking the LLM inflates TPT, and reduces elapsed time, while an LLM refusal should take longer and have a similar TPT to benign responses.

\subsection{Behavioral Monitoring}
\label{behavioral_monitoring}

\begin{table*}[t]
\centering
\small
\renewcommand{\arraystretch}{1.4}
\begin{adjustbox}{max width=\textwidth}
\begin{tabular}{l l l l p{8.5cm}}
\hline
\textbf{Modality} & \textbf{Feature} & \textbf{Test} & \textbf{Direction} & \textbf{Rationale} \\
\hline
Lexical & LLM Judge Block Language & FE & M rate $>$ B rate & LLM judge flags responses containing verbose guardrail-block language such as ``Blocked due to X'', or ``Confidence: xx\%''. \\
Lexical & Per-response prevalence & FE & M rate $>$ B rate & Templated block strings recur across malicious responses, producing the same response far more often than in benign. \\
Timing  & Time-per-token (TPT) & KS & M median $>$ B median & Block paths skip LLM generation and return short canned text, so per-token time is dominated by fixed HTTP latency. \\
Timing  & Elapsed wall-clock & KS & M mean $<$ B mean & Skipping LLM generation sharply reduces total request duration. \\
HTTP    & Status code presence/rate & FE & per-direction & Non-200 codes (e.g. 403, 422) appear when middleware rejects; some success codes can disappear entirely under block. \\
HTTP    & Header presence & FE & per-direction & Security middleware injects block-specific headers, or strips ones present on normal paths. \\
HTTP    & Header value distribution & FE & M top value $>$ B rate & Block paths return distinct values for headers such as \texttt{server} or \texttt{content-type}. \\
HTTP    & Body field flags & FE & per-direction & Response fields appear in block payloads but are absent in normal responses. \\
\hline
\end{tabular}
\end{adjustbox}
\caption{\textbf{Features assessed during behavioral monitoring}. Includes outlined statistical test applied (KS = Kolmogorov–Smirnov; FE = Fisher's Exact 2$\times$2), the direction-of-effect required to indicate a guardrail block, and the rationale for that direction. M = malicious prompt-set, B = benign prompt-set.}
\label{tab:features_to_approaches}
\end{table*}

To successfully detect whether a guardrail has triggered within a target AI system, it is necessary to assess if the state of the system has changed under malicious interactions. Understanding the behavior of the target AI system under both benign and malicious conditions enables us to establish a baseline for normal operation, and subsequently identify deviations that may indicate guardrail activation. To achieve this, we propose the use of a behavioral monitoring system which can be used to monitor the behavior of the target AI system by establishing a baseline, and then assessing how the behavior deviates during interactions that are designed to trigger guardrail blocks.

\textbf{Behavior Monitoring.} The behavior of the target is monitored by sending the benign and malicious prompt-sets to the target, and gathering features discussed in Section \ref{feature_extraction}. Features gathered during the benign prompt-set are used to establish a baseline of normal operating behavior, while features gathered during the malicious prompt-set are used to assess for system state change. Two methods of statistical testing are used, 1) Kolmogorov–Smirnov (KS) Test, and 2) Fisher's Exact (FE) Test. The approach used depends on the type of feature being assessed, such as using the KS Test for continuous features such as per-token latency, and FE Test for categorical features such as status codes (see Table \ref{tab:features_to_approaches} for a full breakdown). Every \textit{p-value} is then adjusted using the Benjamini-Hochberg FDR correction to account for multiple comparisons, and with \textit{q-value} cutoffs of $\alpha$ = 0.05 / 0.01 / 0.001 being used to determine the confidence of a guardrail block signal at weak, moderate, and high respectively. Finally, the direction-of-effect is used to determine whether the change in behavior is indicative of a guardrail block, such as an increase in non-200 status codes or a decrease in response time. 

\newpage
By using the benign prompt-set and a malicious prompt-set, we use the behavioral monitoring approach to measure the confidence whether a guardrail block has occurred within the target AI system. The resulting comparisons for each feature are then used as \textit{signals} to determine the presence of a guardrail block, with the strength of the signal being determined by the confidence of the statistical test, and the number of features that indicate a guardrail block.

\subsection{Guardrail Characteristics Identification}

The resulting signals ascertained from the behavioral monitoring can be used to determine a range of characteristics about the guardrail, such as its \textit{detection capability} and \textit{block pattern}.

\newpage
\textbf{Detection Capability.} In addition to determining if a guardrail exists within a target AI system, the behavioral monitoring approach can also be used to ascertain the characteristics of the guardrail, such as its ability to detect specific categories of malicious content. This is performed by creating multiple malicious prompt-sets, each designed to trigger guardrail blocks in a specific category we are assessing. By capturing q-values across HTTP, Lexical, or Timing features across the different malicious prompt-sets we can determine whether the guardrail is designed to detect specific categories of content and the degree of effectiveness based upon the q-value cutoffs.

\textbf{Block Pattern Fingerprint.} Using the resulting signals from the behavioral monitoring approach, we can begin to create a fingerprint of the guardrail block pattern, recording the required features, capturing the specific patterns that the target system exhibits when a guardrail block occurs. A guardrail block pattern fingerprint is only created for targets that have been identified with high confidence as having guardrails. This fingerprint can then be used to determine if a guardrail block has occurred in any future interactions with the target system, allowing us to determine if a request was blocked by a guardrail or rejected by the underlying LLM, enabling adversaries to optimize attack strategies.

\section{Experiment Setup}
\label{experiment_setup}

\begin{table}[t]
\centering
\small
\renewcommand{\arraystretch}{1.4}
\begin{adjustbox}{max width=\columnwidth}
\begin{tabular}{l p{4.6cm}}
\hline
\textbf{Guardrail} & \textbf{Detection categories} \\
\hline
Protect AI DeBERTa v2 \citep{deberta-v3-base-prompt-injection-v2} & Prompt injection \\
OpenAI Omni Moderation \citep{openai-omni-moderation} & Toxicity \\
DuoGuard 1.5B \citep{deng2025duoguardtwoplayerrldrivenframework} & Jailbreak, Toxicity \\
Meta Prompt Guard v2 \citep{meta-llama-llama-prompt-guard-2-86m} & Jailbreak, Prompt injection \\
LLMGuard* \citep{llm-guard} & Jailbreak, Prompt injection, Toxicity \\
Microsoft Azure Prompt Shield \citep{microsoft-jailbreak-detection} & Jailbreak, Prompt injection \\
Microsoft Azure Content Safety \citep{microsoft-content-safety} & Toxicity \\
Microsoft Foundry Guardrails* \citep{microsoft-foundry-guardrails} & Jailbreak, Prompt injection, Toxicity \\
LlamaFirewall* \citep{meta-llama-firewall} & Jailbreak, Prompt injection \\
\hline
\end{tabular}
\end{adjustbox}
\caption{\textbf{Evaluated guardrails and their detection capabilities.} Vendor-stated coverage is treated as the ground truth when assessing whether the behavioral monitoring approach correctly attributes a block to a guardrail's intended detection capability. * = Guardrail containing multiple scanners configured to detect specific categories.}
\label{tab:guardrails}
\end{table}

\textbf{Targets.} We constructed 6 guardrail block patterns as described in Section \ref{llm_guardrails}, including status code, response body, and header changes (see Appendix Table \ref{tab:block_patterns} for full breakdown). Guardrail block patterns were used across 9 guardrails across a range of classification tasks (see Table \ref{tab:guardrails}). We utilized 3 different LLMs, gpt-4.1, claude-sonnet-4-6, and gemini-2.5-flash, producing 162 unique AI target configurations in total. We include a no-guardrail baseline (i.e. an AI system with no guardrail) to measure the detection false positive rate.

\textbf{System Setup.} All targets were deployed on a Linux Ubuntu 20.04 server with 128GB RAM, an Intel Xeon Gold 6336Y CPU, and two A100 GPUs with 80GB VRAM each. The guardrails were deployed in front of the 3 LLMs, and the guardrail block patterns were configured to trigger when a guardrail block occurred. The guardrails were configured to use the default settings as per the vendor's documentation such as detection thresholds, and suggested scanner configurations. All targets were exposed via an API endpoint only returning the HTTP response to the request, with no additional information about the target system or guardrail being exposed.  

\textbf{Guardrail Reconnaissance Configuration.} We created four prompt-sets each containing 10 prompts: the first being the benign prompt-set, and the other three being malicious prompt-sets targeting specific content categories. The prompt injection and jailbreak sets were sourced from the PI Guard dataset \citep{PIGuard}, while the final set targets a range of safety-violating prompts sourced from the Nvidia Aegis dataset \citep{ghosh-etal-2025-aegis2}. Following the process described in Section \ref{guardrail_detection}, the benign and malicious prompt-sets are sent to each target, gathering the required HTTP, lexical, and timing features. These features are then used to capture guardrail characteristics of the target such as the detection capability and block pattern. After performing behavior monitoring, we then assess the ability for our approach to determine whether a Guardrail Block or an LLM rejection has occurred by sending 100 unseen prompts from the PI Guard dataset, recording the ground truth of what drove the AI system decision.

\textbf{Comparison Baseline.} We evaluated our proposed guardrail reconnaissance methodology against two approaches, 1) \textit{HTTP Recon} which only uses HTTP features to determine the presence of a guardrail, and 2) \textit{LLM Judge} which assesses the entire HTTP response body using an LLM judge to determine guardrail block occurrence. Both approaches were evaluated across the same targets and system setup as our proposed approach.  

\textbf{Metrics.} Throughout the experimentation we use the following metrics to evaluate the performance of the behavioral monitoring approach: 
\begin{itemize}

\item\textit{Guardrail detection accuracy}, measured by correctly identifying the presence of a guardrail in the target AI system under test. 
 \item\textit{Signal strength} of guardrail existence, measured by mean $-\log_{10}(q)$ per signal / channel, with q-value confidence bands for weak (0.05), moderate (0.01), and high (0.001). 
\item\textit{Detection capability}, measured by agreement between recovered categories and vendor-proclaimed categories (see Table \ref{tab:guardrails}). 
\item\textit{Guardrail Block vs LLM rejection classification}, measured by F1 of the block fingerprint on 100 unseen prompts, with ground-truth labels for what drove each decision.
\end{itemize}

\section{Results}
\label{results}

We evaluated our proposed guardrail reconnaissance methodology in three stages, each building on the previous leading towards the objective of distinguishing guardrail blocks from LLM rejection. In stage 1 (Section \ref{guardrail_detection_results}), we assess the capability of our approach against two other guardrail detection methods to determine how effectively we can detect the presence of a guardrail within various target AI system configurations. In stage 2 (Section \ref{guardrail_characteristic_identification_results}), we outline the capability for our approach to determine the detection capability of the identified guardrail, and additionally create a fingerprint of the guardrail block pattern. Finally, in stage 3 (Section \ref{block_vs_rejection_results}), we evaluate the ability for the resulting block pattern fingerprint to distinguish a guardrail block from an LLM rejection on unseen prompts.

\subsection{Guardrail Detection}
\label{guardrail_detection_results}

\begin{figure}[t]
\centering
\includegraphics[width=\columnwidth]{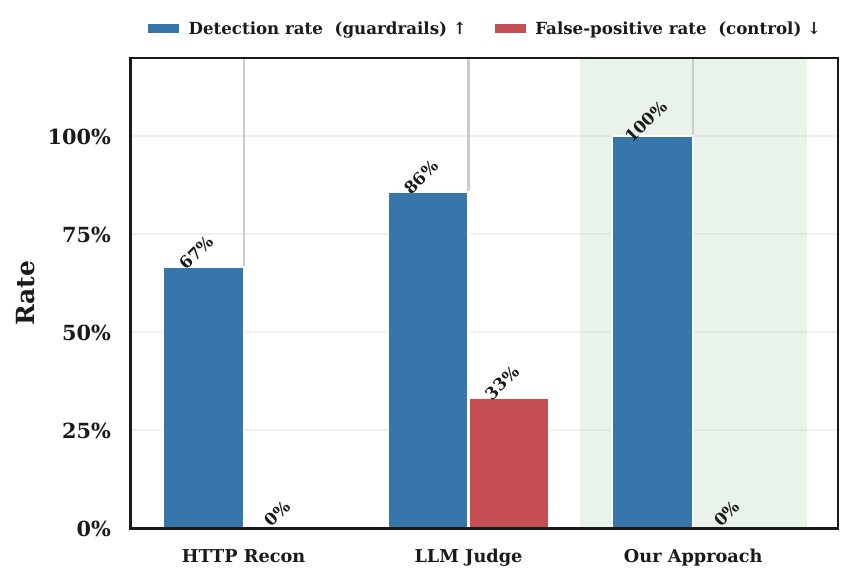}
\caption{\textbf{Guardrail Detection Performance.} Average detection accuracy across all targets including the no-guardrail control.}
\label{fig:benchmark_vs_naive}
\end{figure}

\begin{figure}[t]
\centering
\includegraphics[width=\columnwidth]{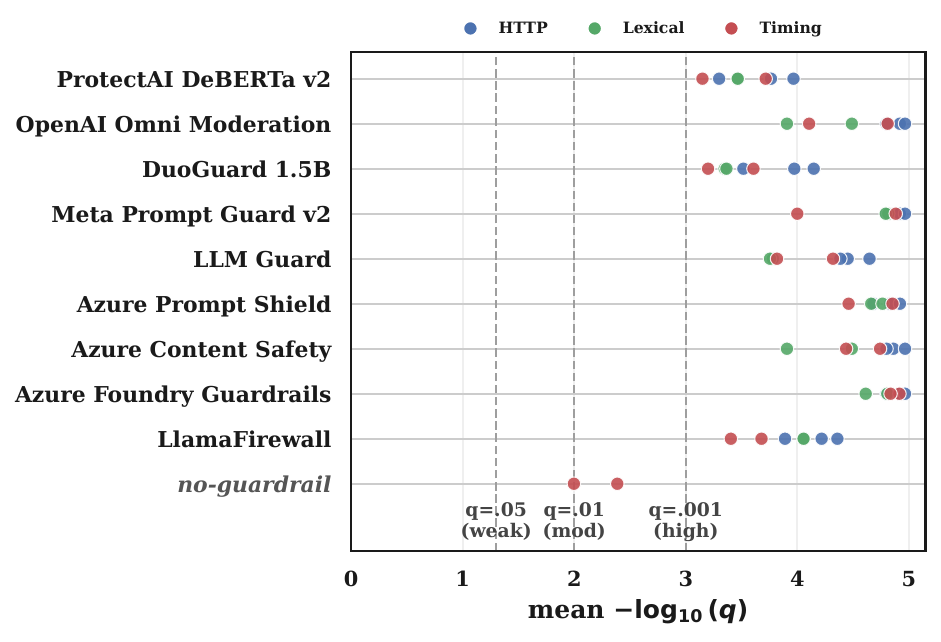}
\caption{\textbf{Guardrail Signal Strength.} Signal strength across all guardrails, 6 block patterns and 3 LLMs. Each marker is one signal, coloured by channel (HTTP, Lexical, Timing), positioned at mean $-\log_{10}(q)$. Higher values indicate stronger statistical separation between malicious and benign prompt-set distributions on signal.}
\label{fig:guardrail_signal_qvalues}
\end{figure}

\begin{figure}[t]
\centering
\includegraphics[width=\columnwidth]{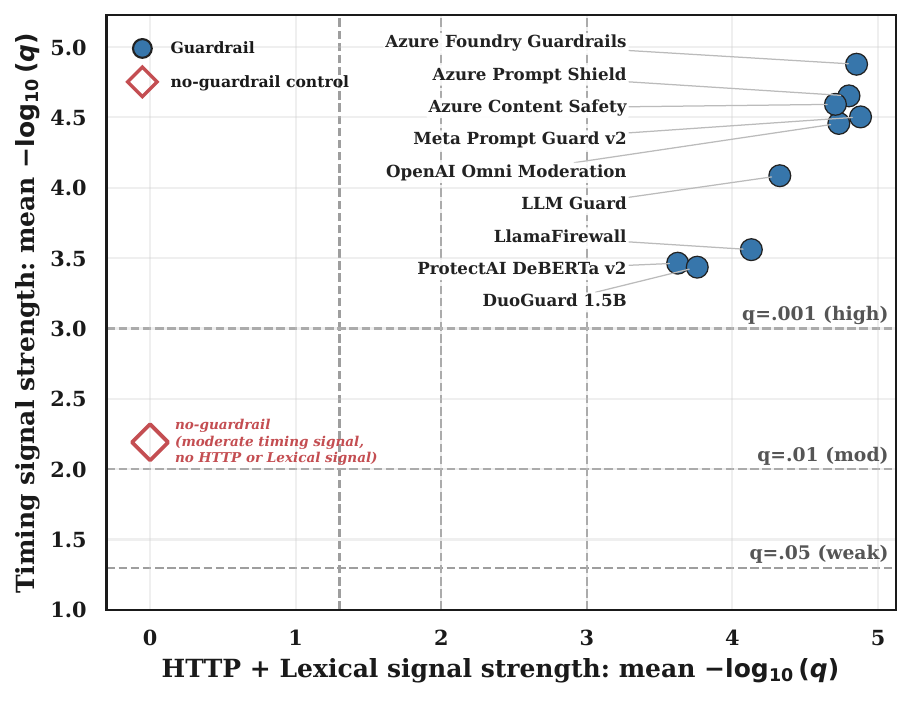}
\caption{\textbf{HTTP \& Lexical Significance.} Each guardrail positioned by its mean HTTP+Lexical signal strength ($x$) against its mean timing signal strength ($y$). Guardrails cluster strongly on both axes, whereas the no-guardrail control collapses to zero HTTP+Lexical signal while retaining only a moderate timing value.}
\label{fig:guardrail_detection_scatter}
\end{figure}

\textbf{Detection Accuracy.} Figure \ref{fig:benchmark_vs_naive} shows the detection accuracy of our approach compared to two approaches \textit{HTTP Recon} and \textit{LLM Judge}. Our approach achieves a 100\% detection accuracy across all targets, including no false positives on the no-guardrail control. In contrast, HTTP Recon achieves 67\% detection accuracy, with 0\% false positives on the no-guardrail control, lacking the ability to detect guardrails with block patterns that do not modify the HTTP layer. LLM Judge achieves 86\% detection accuracy, with a 33\% false positive rate on the no-guardrail control, while the LLM judge is effective, it struggles to distinguish guardrails where the block pattern is not verbose.

\textbf{Feature Detection Strength.} Figure~\ref{fig:guardrail_signal_qvalues} shows the strength of the guardrail block signals across all features. Every guardrail resides within the high-confidence cutoff ($q<0.001$), with the HTTP and Lexical channels producing the strongest and most significant signals. Timing features are also significant across all targets, however they introduce false positives on the no-guardrail control, where timing sits between $q<0.05$ and $q<0.01$ producing a moderate signal (see Appendix Figure \ref{fig:timing_distribution}). Figure~\ref{fig:guardrail_detection_scatter} makes this channel split explicit, where every guardrail sits high on the HTTP+Lexical axis, while the no-guardrail control collapses to zero on that axis. Together these results highlight that HTTP and Lexical features are the most reliable for detecting guardrail blocks, while timing features can be used as a supplementary signal but not directly relied upon due to the noise it can introduce\footnote{We include supplementary figures highlighting signal strengths across all features in the Appendix \ref{sec:supplementary_figures}.}.

\subsection{Guardrail Characteristic Identification}
\label{guardrail_characteristic_identification_results}

For the guardrail characteristic identification, we tune detection thresholds based upon the guardrail detection results, where the HTTP and Lexical channels are used as the primary signals to drive guardrail detection. By evaluating the guardrail detection capability across the 3 malicious prompt-sets, we can determine if the guardrail is designed to detect specific categories of malicious content, and the strength of that detection capability.

\begin{figure*}[t]
\centering
\includegraphics[width=\textwidth]{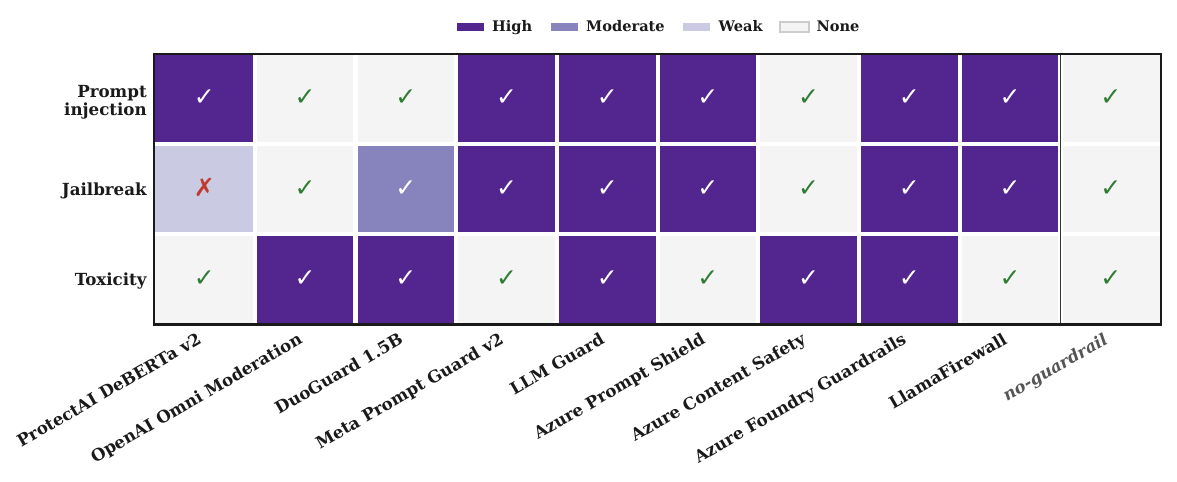}
\caption{\textbf{Guardrail Detection Capability.} Guardrail detection strength across three malicious prompt-sets highlighting suspected content guardrail detects. \cmark{} = Matches vendor-claimed detection capability, \xmark{} = Does not match vendor-claimed detection capability.}
\label{fig:guardrail_dataset_discriminability}
\end{figure*}

\textbf{Detection Capability.} Figure~\ref{fig:guardrail_dataset_discriminability} shows the effectiveness of our approach to detect the content category each guardrail is designed to block across the three malicious prompt-sets. We observed that the approach has strong performance in detecting vendor-claimed detection capabilities for all but one of the evaluated guardrails. Protect AI DeBERTa v2 was the only partial case, where the vendor claims the guardrail is designed to detect only prompt injections, while our approach suggests a weak signal ($q<0.05$) for jailbreaks, indicating weak presence of the target blocking this content. We attribute this to the latent overlap between prompt injection and jailbreak content, which makes the two categories difficult to separate cleanly \citep{hackett-etal-2025-bypassing, bassani2025guardrail}. We observed that the only occurrence of a non-high signal in the correctly identified categories was DuoGuard 1.5B, where jailbreaks were identified with moderate strength. While this is still an adequate signal to indicate DuoGuard detects jailbreaks, it indicates that the jailbreak malicious prompt-set doesn't fully generalise to all guardrails where we would expect a high signal.

\textbf{Block Pattern Fingerprint.} We observed across all target configurations that the generated guardrail block pattern fingerprint was able to capture 100\% of the expected unique patterns of each guardrail block setup (see Appendix Figure \ref{fig:fingerprint_fidelity}). The fingerprint captures the unique patterns of each guardrail block, and can be used to determine if a guardrail block has occurred in future interactions with the target system. This enables us to distinguish between a guardrail block and LLM rejection, allowing adversaries to optimize attack strategies based upon the reason behind the block or rejection.

\subsection{Distinguishing Guardrail Blocks from LLM Rejections}
\label{block_vs_rejection_results}

\begin{figure}[t]
\centering
\includegraphics[width=\columnwidth]{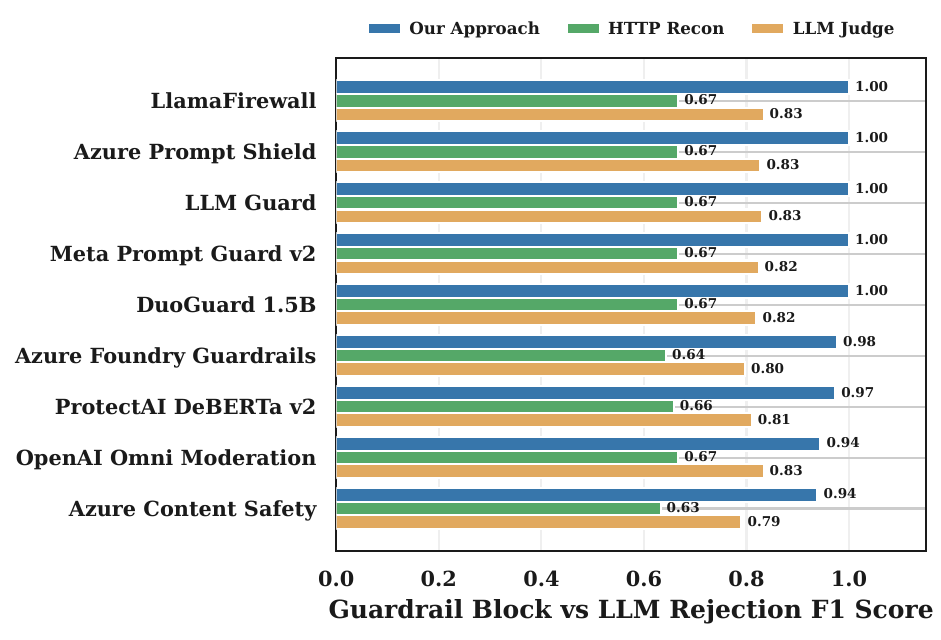}
\caption[Guardrail Block vs LLM Rejection.]{\textbf{Guardrail Block vs LLM Rejection.} F1 of the guardrail block fingerprint at distinguishing a real guardrail block from an LLM rejection on 100 unseen prompts across all targets\protect\footnotemark{}. Higher is better, with $1.00$ being perfect separation.}
\label{fig:block_fingerprint}
\end{figure}
\footnotetext{Guardrail block fingerprinting is only applied to targets with guardrails identified (Section~\ref{guardrail_detection_results}), hence no-guardrail control is excluded.}

Using the guardrail block fingerprint generated from the guardrail characteristic identification step, it is possible to determine whether a guardrail block or an LLM rejection has occurred when using a set of prompts unseen by the target AI system (i.e. a new set of prompts that were not leveraged during adversarial reconnaissance).

\newpage
\textbf{Guardrail Block Classification Performance.} Figure~\ref{fig:block_fingerprint} shows the performance of our approach compared to the two approaches (see Section \ref{experiment_setup}) at distinguishing a guardrail block from an LLM rejection. We only applied fingerprinting once a guardrail has been detected (Section~\ref{guardrail_detection_results}), therefore we restrict this evaluation to guardrail enabled targets and exclude the no-guardrail control. We observed that our approach is highly accurate achieving an average F1 score across all targets of $0.98$, with 5 out of 9 guardrails achieving an average F1 of $1.00$, followed by Microsoft Foundry Guardrails at $0.98$, Protect AI DeBERTa v2 at $0.97$, and with OpenAI Omni Moderation and Microsoft Azure Content Safety both at $0.94$. The two other approaches showed significantly lower performance, with HTTP Recon achieving the lowest average F1 score of $0.66$, followed by LLM judge with $0.81$. All techniques showed varying degrees of success in distinguishing guardrail blocks from LLM rejections, but all had their lowest F1 scores against Microsoft Azure Content Safety. These results demonstrate that our approach is effective at determining if a rejection was caused by a guardrail or the LLM once a guardrail block fingerprint has been generated.

\section{Discussion}
\label{discussion}

\textbf{Benefits to Adversarial ML and Offensive Security Researchers.} Our proposed guardrail reconnaissance approach relaxes a well established threat model across literature in terms of knowledge of guardrail presence and activation. The results found in Section \ref{guardrail_detection_results} demonstrate that only 40 prompts are required to conduct guardrail behavioral analysis. Figure \ref{fig:block_fingerprint} demonstrated high effectiveness in differentiating a guardrail block from an LLM rejection, therefore enabling adversarial researchers to build more effective attack strategies that adapt dependent on the rejection cause, such as using specific adversarial perturbation techniques to evade guardrail detection, as well as jailbreak techniques that bypass LLM safety alignment. Moreover, our approach can be leveraged to perform guardrail reconnaissance in AI red teaming, whereby guardrail detection and fingerprinting can identify the presence of guardrail and its respective capabilities across content categories, enabling AI red teamers to understand guardrail's capabilities and limitations.

\textbf{Guardrails Exhibit Unique Blocking Patterns.} We observe that guardrails exhibit varying degrees of signal strength across the different features, with some guardrails exhibiting strong signals across all channels, while others only exhibit strong signals across a subset of channels. While the effectiveness of the approach relies on the generalizability of the malicious prompt-set to trigger guardrail blocks, the varying signal strengths indicate the possibility of fingerprinting specific guardrails based upon the unique patterns they exhibit. Additionally we note that you require no knowledge of the guardrail or underlying LLM to perform the guardrail reconnaissance, and the approach can be applied to any target AI system, regardless of the underlying LLM or guardrail vendor.

\textbf{Guardrail Capability at a Glance.} A benefit of our approach is the ability to quickly identify the strength of a guardrail's detection capability across a range of content categories. Vendors often claim that a guardrail is designed to detect a specific category of content, yet the true strength of that capability, and whether it extends to other categories, is rarely clear. The approach can therefore be adapted to guardrail evaluation and benchmarking, assessing detection strength across content categories without the need for extensive manual evaluation, enabling more transparent guardrail capabilities for users and researchers.

\section{Related Work}
\label{related_work}

\textbf{Adversarial Reconnaissance.} In traditional cybersecurity, adversarial reconnaissance is the systematic process of gathering intelligence on a target system, including capabilities, defenses, and architecture to tailor an exploit \citep{adv_recon}. This extends to AI security, whereby an attacker assesses an AI target to optimize attack strategies from observed patterns and behavior \citep{guardrail-mismatch,crescendo}. Successful reconnaissance can move an attacker from a strict black-box threat model to a grey-box model \citep{adv_recon}. AI security literature have demonstrated feasible reconnaissance against AI systems, yielding insight into a target's underlying components and capabilities \citep{das2025promptextractionattacksdefenses}.

\textbf{AI Reconnaissance.} Various methods can extract information about the underlying capabilities and components of a target system. For example, there has been considerable work attempting to extract the system prompt of a target, which can reveal its underlying objective and constraints and additionally the tools it has access to, such as database access, web access, and code execution \citep{das2025promptextractionattacksdefenses}. Reconnaissance of a target's components involves identifying the specific models and architectures it uses \citep{llmmap}. This can include identifying the underlying LLM, distinguishing between providers depending on the responses received from the target system \citep{llmmap}.

\textbf{Guardrail Reconnaissance.} In the domain of guardrail reconnaissance, the closest work to ours is a guardrail fingerprinting approach \citep{yang2026peeringshieldguardrailidentification}, which uses an AP-Test to identify which guardrail is deployed within a target system. While identification is performed under black-box access, the approach relies on adversarial probing prompts that must be generated offline in a white-box setting against the candidate guardrails. Crucially, this approach assumes a guardrail is already present and only seeks to identify it, and neither detects whether a guardrail exists nor distinguishes a guardrail block from an LLM rejection. In contrast, our work not only detects the presence of a guardrail under zero knowledge of the target, but also determines its detection capability across a range of content categories and characterizes the block patterns it exhibits. A related guardrail characteristic discovery appears in \citep{zhou2026promptoverflowguardrailinspects}, a guardrail evasion approach that uses a prompt overflow to bypass guardrails by locating where in a prompt the malicious content escapes detection, implicitly revealing the input length up to which the target guardrail is able to inspect.

\textbf{Guardrail Evasion.} There has been extensive work in literature on evading guardrails using various adversarial techniques such as character injection and adversarial ML evasion \citep{guardrail-mismatch, hackett-etal-2025-bypassing, bassani2025guardrail, zhou2026promptoverflowguardrailinspects}. The aim of an adversary is to modify a prompt in such a way that it bypasses the guardrail by exploiting lack of training data, or by exploiting the guardrail's inability to detect certain content categories. While effective, the threat models established in these works assume that the guardrail is known, and that the adversary can determine when a guardrail block has occurred.

Across each area, prior work assumes definitive knowledge of guardrail presence or that its activation is directly observable. Moreover, no existing work distinguishes a guardrail block from an LLM rejection, limiting its applicability to more realistic adversarial emulation, whereby guardrail activation is unknown and commonly conflated with an LLM rejection -- a gap that our work addresses.

\section{Conclusion}
\label{conclusion}

In this paper, we have demonstrated that with as few as 40 prompts, it is possible to detect the presence of a guardrail and its respective behavioral profile at run time. We have shown that by using our proposed approach, we can successfully capture guardrail capabilities to accurately identify guardrail block patterns, and achieve an average F1 score of 98\% when distinguishing whether a guardrail block or an LLM rejection has occurred within any subsequent AI system interaction. Our work highlights the importance of understanding the presence and characteristics of guardrails within AI systems, relaxes well established threat model assumptions within AI security and safety literature, and enables adversarial ML and security researchers to construct more effective attack techniques capable of circumventing AI defenses by adapting based on the detected cause of refusal.

\section*{Acknowledgments}
We would like to thank the Lancaster AI Security team and Mindgard research team for their comments and feedback. This work was supported by the UK Engineering Physcial Science and Research Council (EPSRC) and Mindgard.

\bibliography{paper_references}

\clearpage
\appendix

\onecolumn

\setcounter{figure}{0}
\setcounter{table}{0}
\renewcommand{\thefigure}{\thesection.\arabic{figure}}
\renewcommand{\thetable}{\thesection.\arabic{table}}

\section{Experimental Setup Details}

\begin{table}[H]
\centering
\small
\renewcommand{\arraystretch}{1.4}
\begin{adjustbox}{max width=\textwidth}
\begin{tabular}{l|cc|p{4.6cm}p{2.6cm}|p{5.0cm}p{2.4cm}}
\hline
\textbf{Block pattern} & \multicolumn{2}{c|}{\textbf{HTTP status}} & \multicolumn{2}{c|}{\textbf{Response body}} & \multicolumn{2}{c}{\textbf{Distinguishing headers}} \\
 & Blocked & Allowed & \multicolumn{1}{c}{Blocked} & \multicolumn{1}{c|}{Allowed} & \multicolumn{1}{c}{Blocked} & \multicolumn{1}{c}{Allowed} \\
\hline
no-guardrail (control) & n/a & 200 & n/a & \texttt{\{"response": <LLM output>\}} & --- & --- \\
\hline
verbose-block & 200 & 200 & \texttt{\{"response": "Response blocked due to policy violation. Reason: \ldots. Confidence: \ldots."\}} & \texttt{\{"response": <LLM output>\}} & --- & --- \\
\hline
silent-block & 200 & 200 & \texttt{\{"response": "I'm sorry, I cannot do that."\}} & \texttt{\{"response": <LLM output>\}} & --- & --- \\
\hline
header-warning & 200 & 200 & \texttt{\{"response": "I'm sorry, I cannot do that."\}} & \texttt{\{"response": <LLM output>\}} & \texttt{X-Content-Filter:} \texttt{triggered}\newline \texttt{X-Request-Status:} \texttt{filtered}\newline \texttt{X-Filter-Id:} \texttt{cf-001} & \texttt{X-Content-Filter:} \texttt{passed}\newline \texttt{X-Request-Status:} \texttt{ok} \\
\hline
status-code & 403 & 200 & \texttt{\{"error": "Forbidden", "message": "Your request cannot be processed"\}} & \texttt{\{"response": <LLM output>\}} & --- & --- \\
\hline
error-body & 200 & 200 & \texttt{\{"error": "Invalid request", "message": "Your request cannot be processed"\}} & \texttt{\{"response": <LLM output>\}} & --- & --- \\
\hline
connection-drop & 444 & 200 & empty (zero-byte body) & \texttt{\{"response": <LLM output>\}} & \texttt{Connection: close} & --- \\
\hline
\end{tabular}
\end{adjustbox}
\caption{\textbf{Experiment Targets.} The 6 guardrail block patterns evaluated, plus the no-guardrail control. Each pattern is defined by the HTTP status, response body, and any distinguishing headers returned on blocked and passed requests. ``---'' indicates no distinguishing value.}
\label{tab:block_patterns}
\end{table}

\clearpage
\section{Supplementary Figures}
\label{sec:supplementary_figures}

\begin{figure}[H]
\centering
\begin{subfigure}[t]{0.49\textwidth}
\centering
\includegraphics[width=\textwidth]{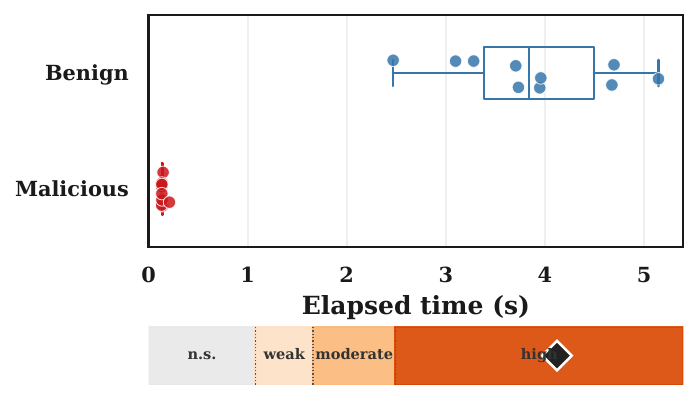}
\caption{Guardrail block (true positive): verbose-block, Meta Prompt Guard v2. The guardrail short-circuits LLM generation, so elapsed times collapse and benign is clearly separated from malicious.}
\label{fig:timing_true_positive}
\end{subfigure}
\hfill
\begin{subfigure}[t]{0.49\textwidth}
\centering
\includegraphics[width=\textwidth]{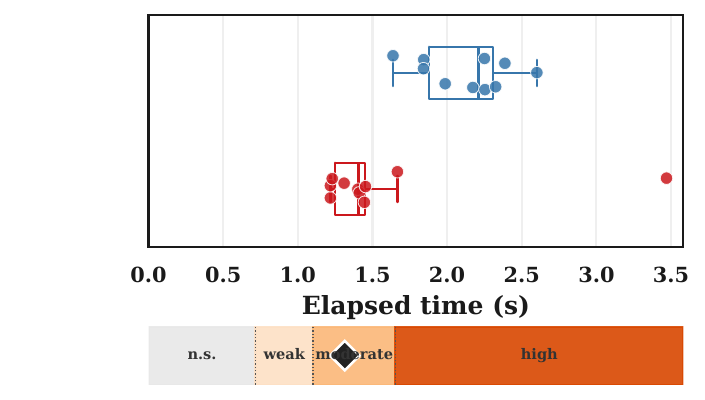}
\caption{No-guardrail noise floor: GPT-4.1, toxicity prompt-set. With no guardrail attached, malicious prompts will intuitively make the LLM generate shorter responses, causing quicker responses.}
\label{fig:timing_noise_floor}
\end{subfigure}
\caption{\textbf{Timing Features False Positive.} The diamond ($\blacklozenge$) marks the experiment's q-value. Both panels reach significance, so timing alone would mistake the no-guardrail refusal in (b) for a guardrail block.}
\label{fig:timing_distribution}
\end{figure}

\begin{figure}[H]
\centering
\includegraphics[width=\textwidth]{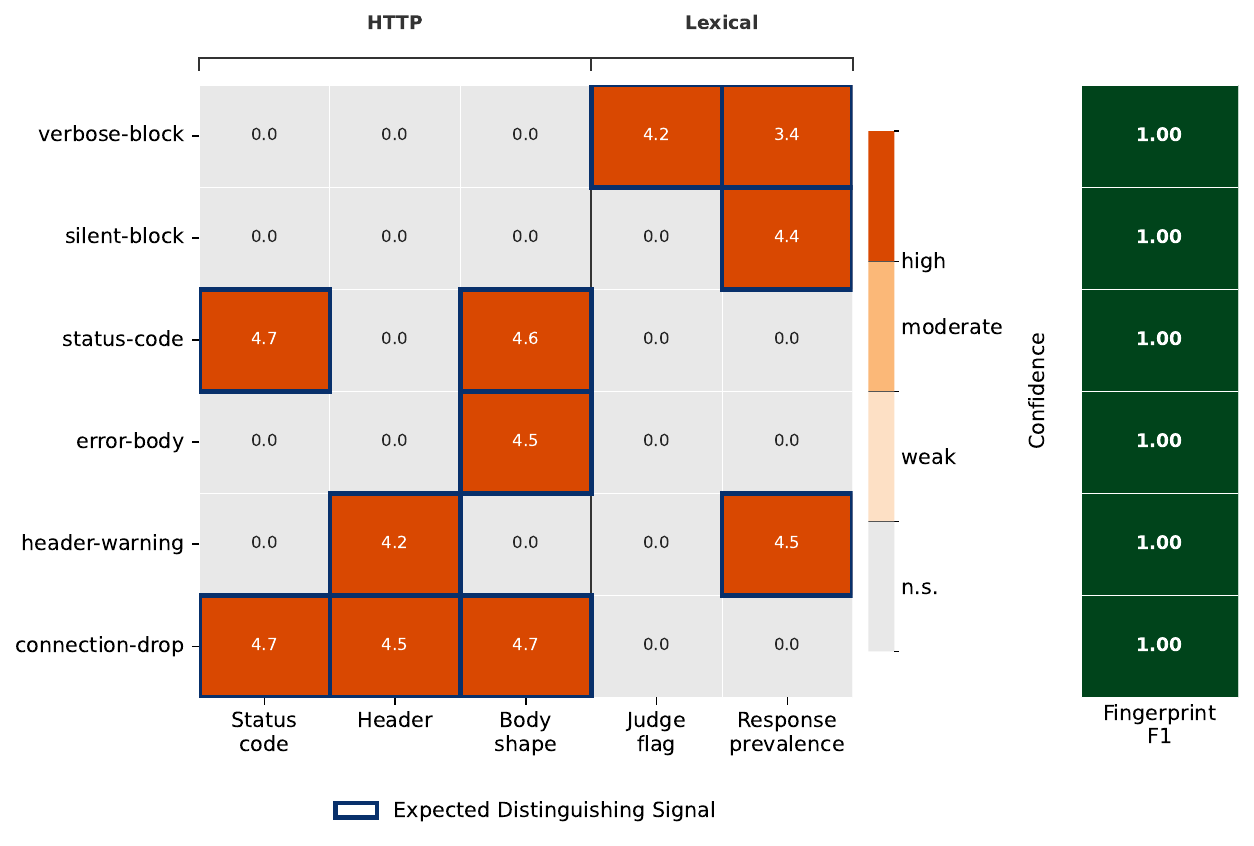}
\caption{\textbf{Recovered block-pattern fingerprints versus documented distinguishing signals.}}
\label{fig:fingerprint_fidelity}
\end{figure}

\begin{figure}[H]
\centering
\includegraphics[width=\textwidth]{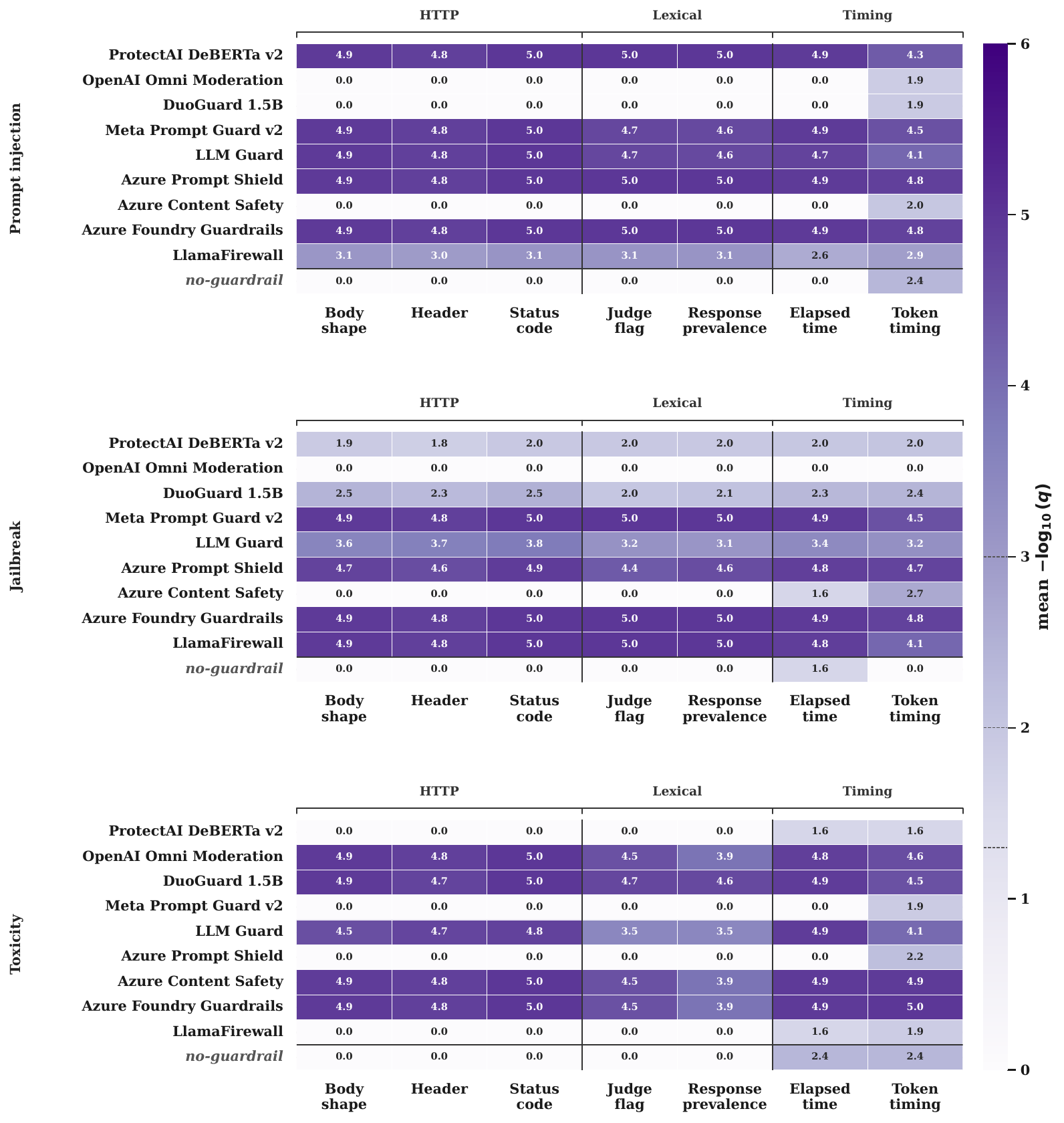}
\caption{\textbf{Per-guardrail signal strength broken down by malicious prompt-set.}}
\label{fig:guardrail_signal_qvalues_per_dataset}
\end{figure}

\begin{figure}[H]
\centering
\includegraphics[width=\textwidth]{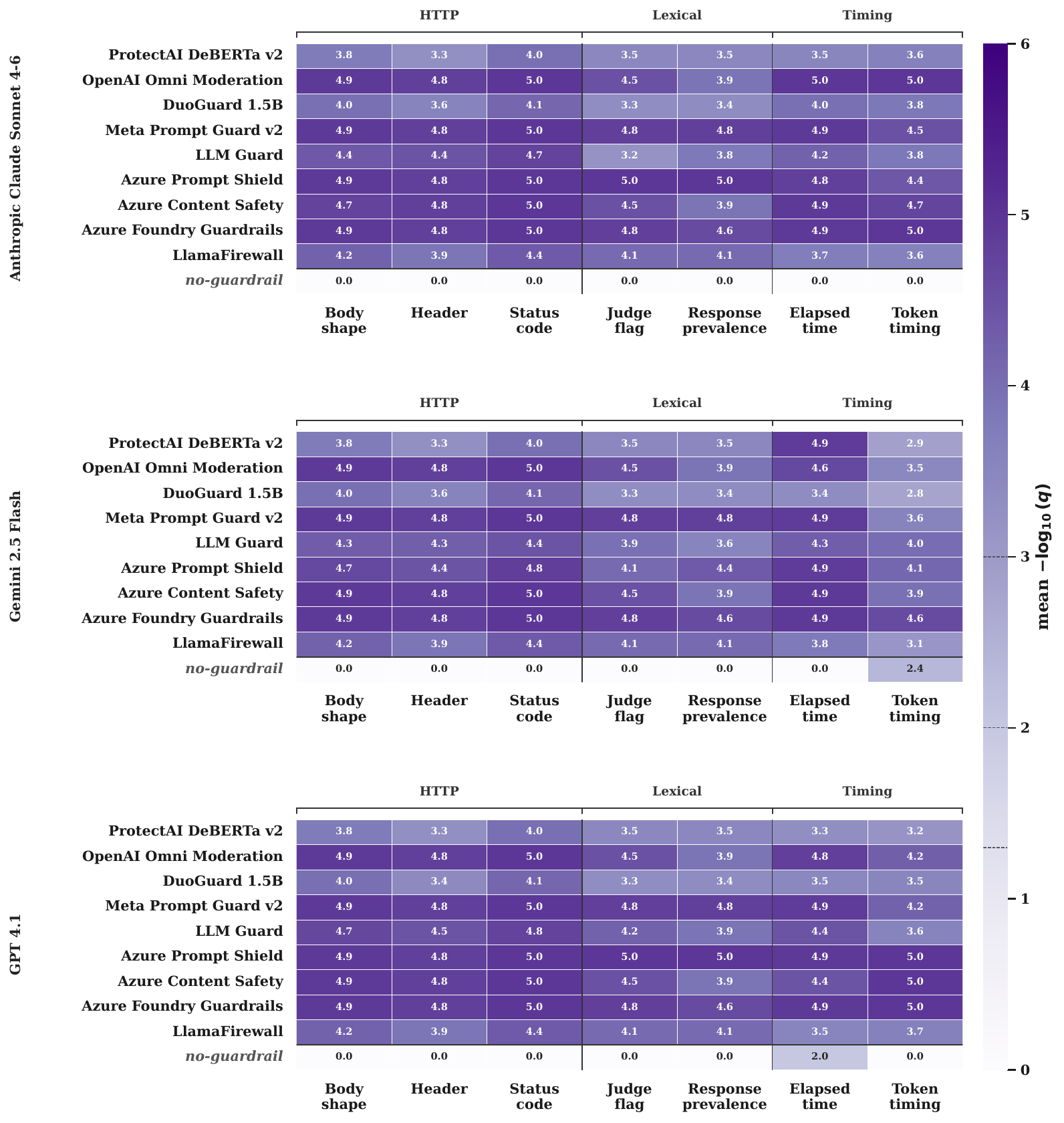}
\caption{\textbf{Per-guardrail signal strength broken down by underlying LLM.}}
\label{fig:guardrail_signal_qvalues_per_model}
\end{figure}

\begin{figure}[H]
\centering
\includegraphics[width=\textwidth]{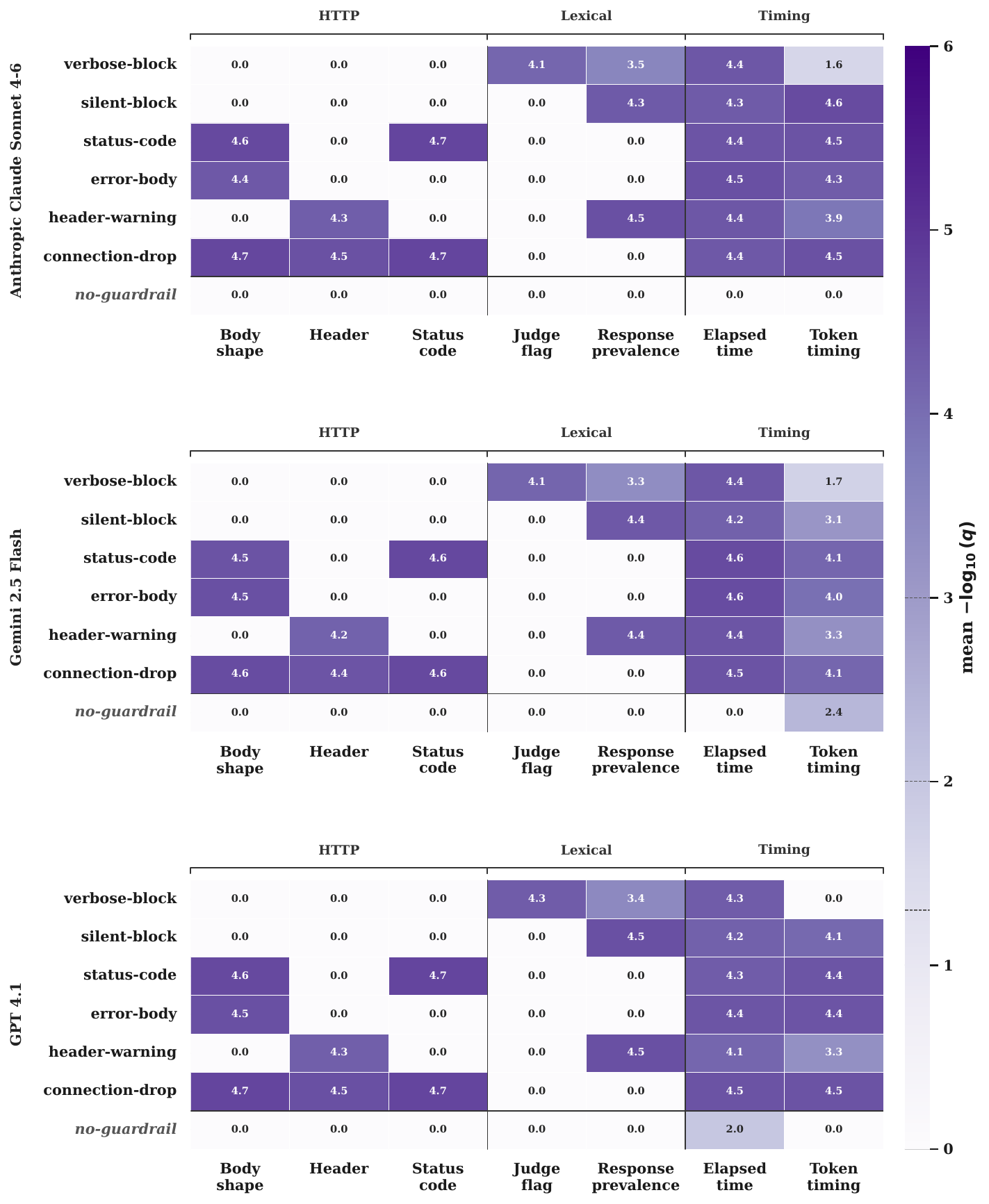}
\caption{\textbf{Per-block-pattern signal strength broken down by underlying LLM.}}
\label{fig:blockpattern_signal_qvalues_per_model}
\end{figure}

\begin{figure}[H]
\centering
\includegraphics[width=\textwidth]{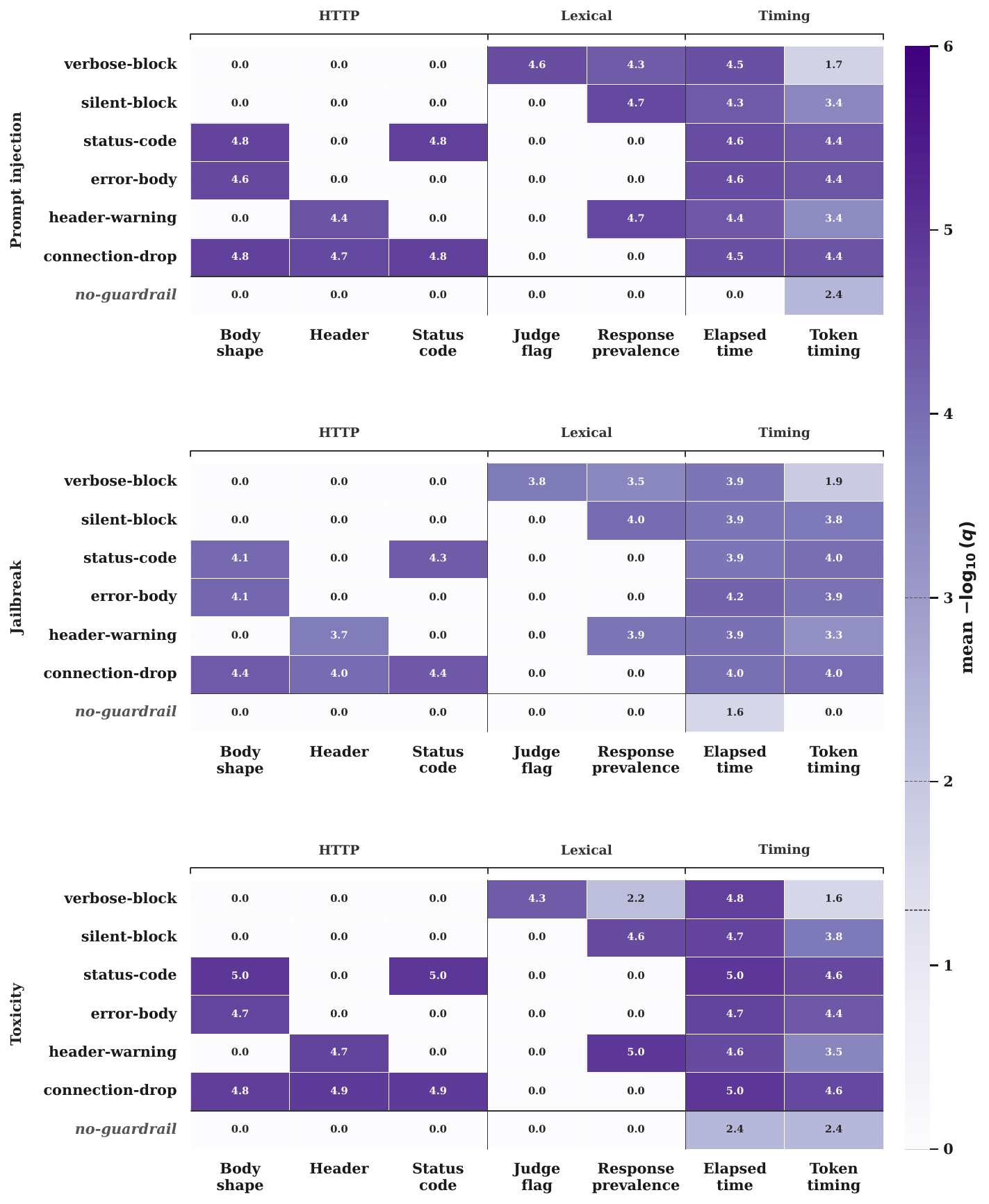}
\caption{\textbf{Per-block-pattern signal strength broken down by each explored malicious prompt-set.}}
\label{fig:blockpattern_signal_qvalues_per_dataset}
\end{figure}

\end{document}